\pdfoutput=1
\documentclass[12pt]{article}
\usepackage{amsthm,amsfonts,amssymb,amsmath,dsfont}
\usepackage[pdftex]{graphicx,graphics,color}
\usepackage[english]{babel}
\usepackage[colorlinks=true,linkcolor=black,citecolor=black,anchorcolor=blue,filecolor=blue,urlcolor=blue]{hyperref}
\usepackage[super,sort&compress]{natbib}
\usepackage[affil-it]{authblk}
\usepackage[T1]{fontenc}
\usepackage[utf8]{inputenc}
\usepackage{lmodern}

\begin{document}

\newcommand{\nn}{\nonumber}
\newcommand{\bra}{\langle}
\newcommand{\ket}{\rangle}
\newcommand{\del}{\partial}
\newcommand{\vt}{\vec}
\newcommand{\dg}{^{\dag}}
\newcommand{\cg}{^{*}}
\newcommand{\T}{^{T}}
\newcommand{\vep}{\varepsilon}
\newcommand{\vth}{\vartheta}
\newcommand{\suml}{\sum\limits}
\newcommand{\prodl}{\prod\limits}
\newcommand{\intl}{\int\limits}
\newcommand{\til}{\tilde}
\newcommand{\wb}{\overline}
\newcommand{\mcl}{\mathcal}
\newcommand{\mfk}{\mathfrak}
\newcommand{\mds}{\mathds}
\newcommand{\mbb}{\mathbb}
\newcommand{\mrm}{\mathrm}
\newcommand{\mnl}{\mathnormal}
\newcommand{\ds}{\displaystyle}
\newcommand{\rmi}{\mathrm{i}}
\newcommand{\rme}{\mathrm{e}}
\newcommand{\rmd}{\mathrm{d}}
\newcommand{\rmD}{\mathrm{D}}
\newcommand{\vphi}{\varphi}
\newcommand{\stm}{\text{\textsf{s}}}
\newcommand{\bth}{\text{\textsf{b}}}
\newcommand{\itr}{\text{\textsf{i}}}
\renewcommand{\mod}{\mathrm{\,mod\,}}
\renewcommand{\b}{\bar}
\renewcommand{\dim}{\mbox{dim}}
\newcommand{\diag}{\mbox{diag}}
\renewcommand{\-}{\,-}
\newcommand{\onlinecite}[1]{\hspace{-1 ex} \nocite{#1}\citenum{#1}} 
\newcommand{\malt}{\mathpzc}
\newcommand{\pd}[2]{\frac{\partial {#1}}{\partial {#2}}}
\newcommand{\pdpd}[2]{\frac{\partial^2 {#1}}{\partial {#2}^2}}
\newcommand{\fd}[2]{\frac{\delta {#1}}{\delta {#2}}}
\newcommand{\pdmix}[3]{\frac{\partial^{2}{#1}}{\partial{#2}\partial{#3}}}
\newcommand{\pdmixxx}[4]{\frac{\partial^{3}{#1}}{\partial{#2}\partial{#3}\partial{#4}}}
\newcommand{\od}[2]{\frac{\dif {#1}}{\dif {#2}}}
\newcommand{\odod}[2]{\frac{\dif^2 {#1}}{\dif {#2}^2}}
\newcommand{\mean}[1]{\langle{#1}\rangle}
\newcommand{\mbf}[1]{\mathbf{#1}}
\newcommand{\itbf}[1]{\textbf{\textit{#1}}}
\newcommand{\sybf}[1]{\boldsymbol{#1}}
\newcommand{\conj}{{*}}
\newcommand{\im}{i}
\newcommand{\ihbar}{\frac{\im}{\hbar}}
\newcommand{\smallfrac}[2]{{\textstyle\frac{#1}{#2}}}

\interfootnotelinepenalty=10000
\renewcommand{\topfraction}{0.85}
\renewcommand{\textfraction}{0.1}
\allowdisplaybreaks[1]
\setlength{\jot}{1ex}
\renewcommand{\theequation}{\thesection.\arabic{equation}}
\numberwithin{equation}{section}
\renewcommand{\thefigure}{\thesection.\arabic{figure}}
\numberwithin{figure}{section}
\renewcommand{\thefootnote}{\roman{footnote}}

\title{Multiscale Approach to Fluid-Solid Interfaces\\
\large An overview of methodologies coupling fluid mechanics
to molecular dynamics and quantum theory}
\author[1]{Thiago F. Viscondi}
\author[1]{Adriano Grigolo}
\author[1]{José A. P. Aranha}
\author[1]{José R. C. Piqueira}
\author[2]{Iberê L. Caldas}
\author[1]{Júlio R. Meneghini}
\affil[1]{Escola Politécnica, University of São Paulo, Brazil}
\affil[2]{Institute of Physics, University of São Paulo, Brazil}
\date{}
\maketitle


\begin{abstract}
In conventional fluid mechanics, the chemical composition and thermodynamic state of a fluid-solid interface are not considered when
establishing velocity-field boundary conditions. As a consequence, fluid simulations are usually not able to generate different outputs 
when interfacial materials are varied. By considering an atomistic description of matter, theoretical determination of material-specific
boundary conditions becomes possible, thereby providing an improved alternative to the completely-invariant no-slip condition. Such a 
scheme constitutes a multiscale approach to fluid dynamics involving essentially two transitions between space-time scales: the first 
concerns the derivation of macroscopic boundary conditions by means of molecular assessment of slip lengths; the second concerns the 
construction of interatomic force fields, required by molecular dynamics simulations, from quantum theory. In this introductory overview 
we discuss some of the fundamental aspects of these problems.

\vspace{0.15cm}

{\bf Keywords:} molecular dynamics, boundary conditions, and multiscale methods.
\end{abstract}


\section{Introduction}
\label{sec:intro}

The purpose of fluid mechanics is to predict the motion of a fluid in a given domain when boundary 
and initial conditions are given. Mathematically, the fluid is regarded as a set of continuum fields 
-- density $\rho$, velocity $\mbf{u}$, and internal energy $\varepsilon$ -- and these must obey balance 
equations that account for overall mass, momentum, and energy conservation. In their most general form, 
the balance equations, given below, are written in terms of the stress tensor $\sybf{\sigma}$ and heat 
flux vector $\mbf{q}$, which describe, respectively, the exchange of momentum and energy between adjacent
fluid elements, thus capturing the effects of dissipation.
\begin{subequations}
\begin{align}
&\pd{\rho}{t} +\nabla\cdot\rho\mbf{u} = 0,
\label{eq:continuity}
\\
&\pd{\rho\mbf{u}}{t} +\nabla \cdot (\rho\mbf{u}\mbf{u} +\sybf{\sigma}) = \rho\mbf{g},
\label{eq:momentum.balance}
\\
&\pd{\rho\varepsilon}{t} +\nabla \cdot (\rho\varepsilon\mbf{u} +\mbf{q}) =
-\sum_{ij} \sigma_{ij} \pd{u_i}{x_j}.
\label{eq:energy.balance}
\end{align}
\end{subequations}
In order to close the system of equations, the continuum formulation requires constitutive
relations: phenomenological laws that connect $\sybf{\sigma}$ and $\mbf{q}$ to the basic field
variables. Most liquids and gases behave as Newtonian fluids and obey the well-known constitutive
laws\cite{landau1987fluid}
\begin{subequations}
\begin{align}
&\sigma_{ij} = (p -\eta\nabla\cdot\mbf{u}) \delta_{ij}
-\mu \left(\pd{u_i}{x_j} +\pd{u_j}{x_i} -\frac{2}{3} \nabla\cdot\mbf{u}\:\delta_{ij}\right),
\label{eq:strain.stress.newtonian}
\\
&\mbf{q} = -\kappa \nabla T.
\label{eq:fourier.law}
\end{align}
\end{subequations}
At this level of theory, different substances are therefore characterized by their transport
coefficients: shear viscosity $\mu$, volume viscosity $\eta$, and thermal conductivity $\kappa$.
Additionally, an equation of  state relating pressure $p$, temperature $T$, and density $\rho$
must be supplied, also allowing $\varepsilon$ to be expressed in terms of these quantities.

On the other hand, from a microscopic perspective, the forces experienced by a fluid element,
the work done on it, and the amount of heat it absorbs or gives are a result of momentum
and energy exchange with the surrounding elements due to molecular diffusion and
collisions. Therefore knowledge of the local molecular state of the fluid, in statistical
terms, should suffice to specify both $\sybf{\sigma}$ and $\mbf{q}$ at each point in space.
For example, in the case of rarefied gases, the Newtonian constitutive laws emerge from 
the Boltzmann equation, the fundamental equation of physical kinetics, when non-equilibrium 
fluctuations are accounted for in the balance equations by means of the Chapman-Enskog and 
other related procedures.\cite{grad1949kinetic, Chapman1990, Kremer2005} In the case of 
liquids, formal expressions for the stress tensor and heat current density were derived 
from statistical-mechanical considerations in a seminal paper by Irving and 
Kirkwood,\cite{irving1950statistical} and are routinely used in the microscopic 
determination of transport coefficients.

In addition to constitutive relations, an appropriate choice of \textit{boundary conditions} at the
fluid-solid interface must be supplied so that the problem expressed in
equations \eqref{eq:continuity}--\eqref{eq:energy.balance} is mathematically well posed. Despite
the manifest importance of interfacial boundary conditions, they can not be fully determined
within the theoretical framework of macroscopic fluid mechanics. Instead their specification
usually relies on empirical observation -- in the case of the velocity field of a viscous fluid,
the condition at the contact surface between a fluid and a fixed solid wall can be suitably
approximated by the \textit{no-slip condition}, which states, on the grounds of vast experimental
evidence, that the flow velocity vanishes at the interface.

Again, from a more fundamental standpoint, the actual condition to be imposed on the flow variables
at the fluid-solid interface would be more naturally formulated in terms of the momentum and energy
exchange between adjacent fluid and wall elements, being thus determined by the molecular state of
the contact region. The information required for specifying this type of boundary condition is
out of reach of continuum theories -- it can only be accessed if the structure of the
interface is described at a microscopic level.

It then becomes clear, from the above considerations, that direct microscopic-continuum
coupling schemes might be devised where specific constitutive rules and boundary conditions for a
given substance are obtained by means of atomistic simulation methods\cite{Asproulis2012, Shang2012a,
Borg2013a, Cosden2013, Markesteijn2014} such as \textit{molecular dynamics}, a term which comprises
a set of computer-simulation tools capable of assessing physical properties of materials by performing
statistical evaluations on the trajectory of a representative many-molecule system governed by classical
equations of motion.\cite{allen2017computer}

In molecular dynamics simulations, molecules are contained in a finite domain -- the
simulation box -- and interact according to specially designed \textit{force fields},
the latter usually conveyed through an effective potential energy function. Additionally,
the equations of motion can be modified so as to make the system's trajectory sample
specific thermodynamical ensembles.\cite{hoover1985canonical, hoover1986constant}
Several fluid properties can be reliably obtained from standard equilibrium molecular
dynamics: specific heats, equations of state, and even some transport coefficients,
such as viscosity and diffusivity;\cite{rapaport2004art} but there are also ways
to enforce non-equilibrium constraints in the simulation box (e.g.~strain rate or
temperature gradients) and to efficiently measure the material's response -- such
techniques can be employed, for instance, to determine strain-stress relations
in fluids and detect deviations from the Newtonian laws expressed in equations
\eqref{eq:strain.stress.newtonian} and \eqref{eq:fourier.law}.\cite{edberg1987rheology,
morriss2007statistical} Similarly, violations of the no-slip condition can be quantified
by considering a microscopic setup of the fluid-solid interface where fluid molecules
participate in a shear flow.

This type of approach, once complemented with protocols that correctly communicate the
information across the micro and macro scales, enables a more detailed fluid-dynamical description,
one which is expected to be important not only in the case of unconventional fluids, but also when
critical phenomena takes place (e.g.~shock waves, phase transitions) in otherwise Newtonian fluids,
or when modeling particular regions of the flow such as fluid-fluid and fluid-solid interfaces.

The present contribution provides an overview of the rich physics encountered in such
a multiscale framework, taking the boundary-condition specification problem as a guideline.
We begin, in section~\ref{sec:material.specific}, by examining the continuum-molecular link:
the key concept of \textit{slip length} is introduced and the issues related to its evaluation
on a fluid-solid interface are discussed, following a brief account of numerical studies on the
subject. Then, in section~\ref{sec:force.fields}, we explore the quantum-mechanical origin of
the force fields utilized in molecular simulations, and describe in broad terms the strategies
employed for their construction. Final remarks are made in section~\ref{sec:final.remarks}.



\section{Material-specific boundary conditions at fluid-solid interfaces}
\label{sec:material.specific}

The investigation of viscous flows past solid surfaces is a central concern of fluid mechanics,
as this fundamental issue is ubiquitous in engineering problems. In numerous situations of
practical interest the equations governing the flow are solved using the no-slip condition,
which asserts that the tangential velocity of the fluid, in the rest frame of the solid surface,
is zero at their interface.\cite{day1990no, lauga2007microfluidics} Although results obtained
under this assumption prove to be consistent with most physical observations, one must recognize
that the no-slip condition is intrinsically limited, since it essentially states that the shear
stress applied by \textit{any} solid surface on \textit{any} fluid is \textit{always} sufficient
to bring the flow at the interface to rest. Note that there is no question about the normal component
of fluid velocity at the interface, since its value, by the very definition of a flow-constraining
solid surface, must be invariably taken as zero. The following discussion then falls only on the
tangential velocity of the flow at the fluid-solid interface.

As one might expect, the underlying limitations of the no-slip condition do lead to unphysical
or incorrect results in several cases of scientific and technological significance.\cite{granick2003slippery,
ellis2004slip, neto2005boundary} For example, the no-slip condition causes the divergence of physical quantities,
such as the stress tensor and the energy dissipation rate, when trying to describe the contact line of a two-phase
fluid moving along a solid surface.\cite{thompson1989simulations, koplik1998no} The no-slip condition also fails
to correctly quantify the velocity fields of internal flows through diminutive channels, such as nanotubes, porous
media, membranes, and microfluidic devices.\cite{sokhan2001fluid, sokhan2002fluid} In these systems, surface effects
become dominant over the bulk behavior of the fluid, often leading to novel phenomena.

Furthermore, the no-slip condition is \textit{material invariant}, that is, its prescribed
interfacial velocity does not depend on the molecular composition, spatial structure, and 
thermodynamic conditions of the fluid or the solid surface. For instance, as a consequence 
of the no-slip condition, the same velocity field would be predicted for the flow of water 
over hydrophobic and hydrophilic surfaces. The no-slip condition also does not take into 
consideration whether the fluid is a liquid or gas, whether the solid surface is remarkably 
rough or perfectly smooth, whether the interface is at an extremely high or particularly 
low temperature, and so on.

Determination of material-specific boundary conditions can be achieved with the aid of molecular
dynamics. In this case, the velocity field at a fluid-solid interface is not preliminarily
specified, but obtained as a statistical result arising from the collective behavior of many
interacting atoms. That is, by considering a suitable choice of interaction forces between a
relatively large number of particles, molecular dynamics simulations are able to accurately
determine physical properties of fluids, solids, and their interfaces. Once obtained,
material-specific boundary conditions can guide the development or selection of materials with
desired interfacial characteristics for particular applications. For example, an accurate interface
modeling could assist in the design of internal coatings capable of reducing the head loss in ducts
carrying gas or oil.

\subsection{Slip boundary condition}

All issues associated with the no-slip condition, including its material invariance, can be addressed
by introducing the concept of \textit{fluid slip}, that is, by allowing a mismatch between the tangential
velocities of the fluid and the solid surface at their interface. In mathematical terms, this procedure
amounts to replacing the no-slip condition, which constitutes a Dirichlet or first-type boundary condition,
by the following relation:
\begin{equation}
	u_{t}=L_{s}\frac{\del u_{t}}{\del n},
	\label{eq:SlipLength}
\end{equation}
\noindent which represents a Robin or third-type boundary condition for the fluid velocity field.
In the above equation, $u_{t}$ symbolizes the tangential component of the interfacial fluid velocity,
considering the solid surface at rest, and $\del u_{t}/\del n$ denotes the shear rate at the interface
or, equivalently, the derivative of the tangential velocity with respect to the surface normal direction.
The parameter~$L_{s}$, for which identity~\eqref{eq:SlipLength} establishes a definition, is known as
\textit{slip length}.

\begin{figure}[htb]
	\centering
	\includegraphics[width=0.6\textwidth]{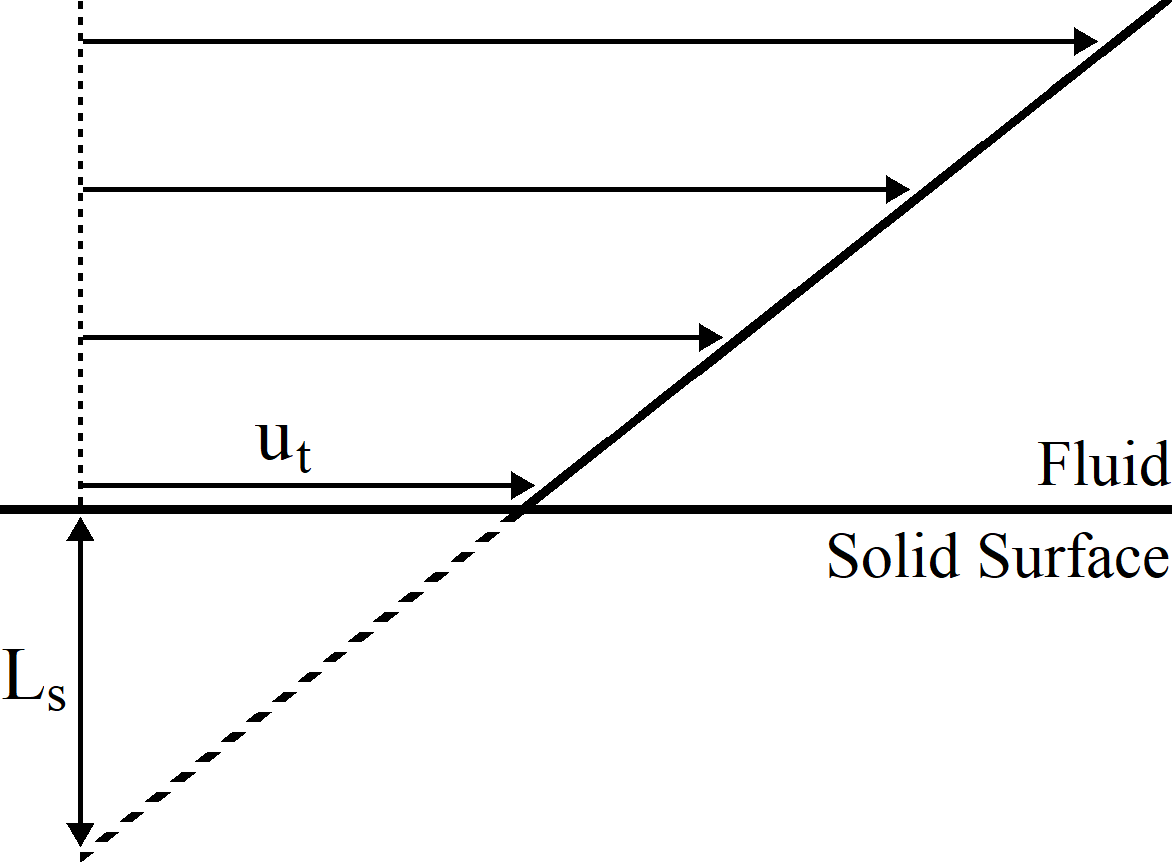}
	\caption{Depiction of the slip length~$L_{s}$ for a fluid flow at a stationary solid surface.}
	\label{fig:SlipLength}
\end{figure}

According to equation~\eqref{eq:SlipLength}, the slip length can be interpreted
as the distance from the interface by which the fluid velocity field needs to be
linearly extrapolated in order to achieve the velocity of the solid surface. This
geometric interpretation is depicted in figure~\ref{fig:SlipLength}.

Note that, by setting $L_{s}=0$ in identity~\eqref{eq:SlipLength}, the no-slip
condition is recovered. On the other hand, when $L_{s}$ goes to infinity, perfect
slip is obtained. In general, the slip length establishes a proper quantification
for the magnitude of fluid slip over a solid surface. Moreover, the slip-length
value can incorporate information about the materials composing the fluid-solid
interface, such as roughness and chemical constituents, and also details about
the flow itself, such as local shear rate and thermodynamic conditions.

\begin{figure}[htb]
	\centering
	\includegraphics[width=0.65\textwidth]{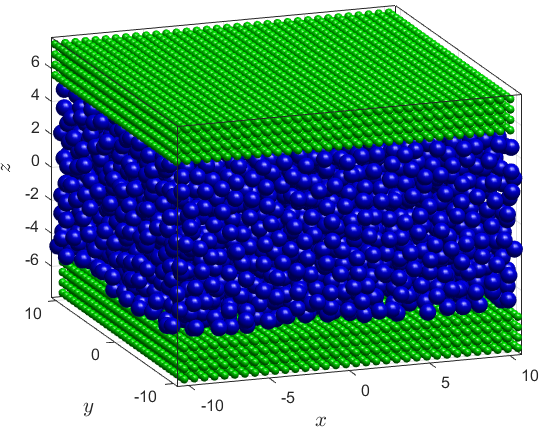}
	\caption{Snapshot of a molecular dynamics simulation depicting
	a monatomic fluid (blue spheres) between two rigid solid walls
	(green spheres).}
	\label{fig:CouetteSimulation}
\end{figure}

The slip length, for a particular choice of interfacial materials and their thermodynamic states,
can be evaluated by a direct application of equation~\eqref{eq:SlipLength} to a molecular dynamics
simulation of a simple flow configuration, such as a planar \textit{Couette}\footnote{Planar Couette 
flow is the drag-induced flow of a viscous fluid confined between two parallel-moving flat plates.} or
a \textit{Poiseuille\footnote{Poiseuille flow is the pressure-induced flow of a viscous fluid confined
in a straight duct.} flow}, as portrayed by figure~\ref{fig:CouetteSimulation}. The procedure of extracting 
the slip-length value from a molecular dynamics simulation begins by performing local time averages over 
the particle velocities in a steady flow regime. As a result, the fluid velocity field is obtained, thus 
making readily available the interfacial values of fluid velocity and shear rate, which are then used to 
calculate the slip length according to its definition.

\subsection{Practical considerations on the slip length}

Once the slip-length value is determined by an atomistic simulation,
identity~\eqref{eq:SlipLength} can then be employed in the macroscopic
context, this time as a third-type boundary condition. However, three 
subtle issues must be considered before transitioning from the molecular 
domain to a continuum description of viscous flows.

First, in order to enable the slip boundary condition to be used in
an arbitrary macroscopic problem, regardless of its interface geometry,
the slip length must not depend on the flow configuration in which its
value is evaluated. That is, for a single choice of interfacial materials
and thermodynamic conditions, molecular dynamics simulations of, for example,
a Couette or a Poiseuille flow must provide the same slip-length value. This
is indeed the case, as shown in published works.\cite{koplik1989molecular,
cieplak2001boundary}

Second, in translating an atomistically-evaluated slip length into a macroscopic boundary condition,
the behaviour of this physical quantity as a function of the system size must be known, otherwise
the results of molecular dynamics can not be expected to remain valid on a completely different
spatial scale. In extreme-confinement scenarios, it has been demonstrated that the fluid-slip
magnitude can vary substantially with the size of the channel.\cite{gupta1997shear, jabbarzadeh1999wall,
liu2011validity} As previously mentioned, in situations of molecular-size confinement, surface effects
may become predominant and, as a consequence, evaluated physical properties may differ considerably
from their bulk values. On the other hand, for molecular dynamics simulations of sufficiently large
size, the slip-length value becomes insensitive to further increases in the pertinent system dimensions,
indicating that the interfacial slip is no longer directly influenced by the central part of the fluid
flow, which presumably has already reached the bulk regime. In this size-invariant case, the slip length
is properly characterized as an \textit{interface property}\footnote{That is, a property whose value is
determined only by the variables characterizing the structure and dynamics of the fluid-solid interface
and, therefore, does not depend on the state of the fluid flow outside the immediate vicinity of the
solid surface.} and its value can be readily employed on the macroscopic scale.\cite{xu2007boundary,
thomas2008reassessing, ramos2016hydrodynamic}

The third concern in using equation~\eqref{eq:SlipLength} as a boundary condition is the very fact
that the slip length depends on a large number of parameters describing the fluid-solid interface,
including physical quantities that correspond to dynamical field variables in the macroscopic context.
Therefore, at different interface points and time instants, distinct slip-length values may be supplied
to condition~\eqref{eq:SlipLength}, depending on the local and instantaneous values assumed by the fluid
field variables, such as temperature, pressure, and velocity. There are basically two ways of dealing with
this issue. First, the slip length could be evaluated \textit{in advance} as a function of all its parameters,
considering a sufficiently wide range of their values. In this case, the required number of molecular dynamics
simulations would increase exponentially with the considered number of slip-length parameters. However, this
costly procedure would be performed only once for each choice of interfacial materials and its results could
be promptly employed in an arbitrary number of macroscopic calculations. As a second option, molecular dynamics
simulations could be performed \textit{concurrently} with macroscopic computations. In this situation, the slip
length would be evaluated on the fly, only for the immediately necessary values of its parameters. This method,
in comparison with a complete preliminary mapping of the slip-length parametric dependence, would substantially
reduce the initial investment of computational resources, while the cost of each multiscale simulation, by
itself, would increase.

As consistently reported in the literature,\cite{thompson1997general, priezjev2004molecular, pahlavan2011effect} 
by employing molecular-dynamics simulations, the typical behavior of the slip length as a function of an interface 
variable can usually be identified. In this case, as a result of a regression analysis, an adjustable curve, generally 
specified by a small set of parameters, is used to properly describe the functional dependence of the slip length on 
the examined variable over a fairly wide domain. Once a suitable regression function is found for a particular choice 
of interfacial materials, the computational cost of repeatedly calculating slip-length values during a multiscale simulation 
can be substantially reduced, as only a small number of simulations is required to determine the few adjustable parameters 
characterizing the relation between the slip length and the interface dynamical variable.

\subsection{Aspects determining slip length}

Originally, Navier introduced identity~\eqref{eq:SlipLength} as a constitutive relation, establishing
a linear correspondence between the interfacial values of velocity and shear rate.\cite{navier1823memoire}
For this reason, in the case where slip length is independent of shear rate, equation~\eqref{eq:SlipLength}
is known as \textit{Navier boundary condition}. As extensively demonstrated for polymers melts with the
aid of molecular dynamics simulations,\cite{khare1996rheology, jabbarzadeh2002effect, priezjev2004molecular,
niavarani2008rheological, niavarani2008slip, priezjev2009shear, priezjev2010relationship, priezjev2012interfacial,
priezjev2012fluid} the slip length associated with a non-Newtonian fluid does not remain constant as a function 
of shear rate. This expected result provides an example of violation of the Navier condition, that is, a situation 
in which identity~\eqref{eq:SlipLength} constitutes a nonlinear relation between velocity and shear rate at the 
interface.

Surprisingly, molecular dynamics simulations also indicate that slip length is not shear-rate invariant
even for Newtonian fluids.\cite{thompson1997general, priezjev2007rate} More precisely, it has been shown
that the slip length asymptotically approaches a constant value for decreasing interfacial shear rate,
thus approximately satisfying the Navier condition in a low-shear-rate regime. On the other hand, the
slip length may also increase indefinitely as the shear rate converges to a critical value. Subsequent
work has suggested that slip-length divergence is in fact caused by modelling solid surfaces as completely
rigid molecular structures, since similar behaviour had not been observed for flexible walls.\cite{martini2008slip}
However, further research has shown that a simultaneous rise in the fluid equilibrium temperature, as the shear
rate is increased, had been actually responsible for suppressing unbounded values of slip length.\cite{pahlavan2011effect}
This result illustrates that, in addition to properly describing the interplay between fluid slip and surface
stiffness,\cite{asproulis2010boundary, asproulis2011wall} appropriate choices of energy-dissipation and thermalisation
mechanisms, the so-called \textit{molecular-dynamics thermostats},\cite{hunenberger2005thermostat} are also essential
in determining realistic slip-length values, which would adequately represent an empirical situation.

Material-specific boundary conditions are obtained by an accurate representation,
from a statistical-physics viewpoint, of the substances composing a fluid-solid
interface. In a molecular dynamics simulation, the description of materials is
primarily determined by the choice of effective interaction potentials. This basic
statement compels the investigation of the slip length through variations of the
interatomic force fields.\cite{liu2009flow, liu2010surface, semiromi2010nanoscale,
sofos2013parameters} Another important factor affecting the slip-length value
is the spatial arrangement of the solid-surface molecules, as discussed in
several studies examining the effects of surface patterning and roughness
on fluid slip.\cite{jabbarzadeh2000effect, cottin2003low, cottin2004dynamics,
galea2004molecular, priezjev2005slip, priezjev2006influence, priezjev2007effect,
niavarani2010modeling} The third element influencing slip length, as previously
implied, is the thermodynamic state of the interface. Both temperature and
pressure have a significant impact on the slip-length value, as demonstrated
by molecular dynamics simulations.\cite{guo2005temperature, servantie2008temperature, 
bao2017effects} For instance, as shown for simple fluids,\cite{pahlavan2011effect} 
the slip length presents a linear decreasing behaviour for increasing temperature 
and constant pressure, whereas its value decreases nonlinearly with increasing 
pressure and constant temperature.

In view of the many different aspects directly involved in evaluating slip length from an atomistic perspective, several
efforts have been directed towards understanding the physical mechanisms underlying interfacial slip.\cite{lichter2004mechanisms,
lichter2007liquid, martini2008molecular, yong2010investigating, sochi2011slip, yong2013slip} Among these studies, special attention
has been given to potential correlations between fluid slip and \textit{wettability}.\cite{barrat1999large, barrat1999influence,
nagayama2004effects, huang2008water, sendner2009interfacial, huang2012friction, ramos2016wettability, yen2016effective} As 
expected, molecular dynamics simulations indicate the existence of a quasi-universal relationship, which states that a decrease 
in wettability is consistently accompanied by an increase in the slip-length value, although exceptions to this rule have also 
been found.\cite{voronov2006boundary, voronov2007slip, voronov2008review, ho2011liquid}

Just as the partial differential equations governing the macroscopic dynamics of viscous fluids must be supplied with interfacial
boundary conditions, which need to be evaluated by empirical observations or by an independent theoretical framework acting on a
smaller space-time scale, molecular dynamics simulations must also receive an external input, namely, the classical force fields
determining the effective interaction among atoms. The next section examines the intricate problem of obtaining molecular-dynamics 
force fields from the fundamental theory of matter, \textit{quantum mechanics}.


\section{Force fields in molecular dynamics}
\label{sec:force.fields}

Running a molecular dynamics trajectory requires knowledge of the forces experienced by each molecule inside 
the simulation box. These forces act, by construction, on specific sites of a molecule, most often coinciding 
with nuclear positions, though other possible locations include bond midpoints or, when dealing with large 
biomolecules, coarse-grained atomic clusters. Since forces are repeatedly evaluated at every time step of 
the simulation, computational efficiency demands that they be provided in explicit form. Hence, they are 
usually derivable from a potential energy $E$, a known function of $M$ `atomic positions', hereof meaning 
`generic force centers', belonging to all molecules of the system, the force on atom $A$ obtained through:
\begin{equation}\label{eq.force.as.gradient}
\mbf{F}_A = -\pd{E(\mbf{R}_1, \dots, \mbf{R}_M)}{\mbf{R}_A}
\end{equation}
Thus, in order to build a sensible force field, one searches for an all-atom energy expression capable of correctly
describing the molecular interaction mechanisms most relevant for the substance at hand and which best suits the
thermodynamic conditions at play. Despite the existence of many types of force fields, specifically designed for
gases, liquids, metals, proteins, polymers, and so on, \cite{brooks1983charmm, daw1984embedded, van2001reaxff,
ponder2003force, lopes2015current} the pursuit of new, more accurate, more efficient, and more versatile models
is a never-ending research topic in computational physics and chemistry. Here, we give a brief account of how
force fields can be constructed.

\subsection{Quantum description of a single molecule}
\label{sub:single.molecule}

The energy function in equation~\eqref{eq.force.as.gradient} depends on nuclear coordinates only,
implying that all information related to the distribution of electronic charge is somehow incorporated
into its functional form. To understand how this comes about we start by considering the quantum description
of a single molecule,\cite{atkins2011molecular, szabo2012modern} regarded as a collection of $N_e$ electrons
and $N_n$ nuclei interacting through electrostatic forces.\footnote{In practice only valence electrons need
to be considered -- core electrons and nuclei are combined into effective charge centers with the help of
so-called \textit{pseudopotentials}.} The ensuing theory rests on a crucial observation: since nuclei are 
much heavier than electrons, the time scale for nuclear motion is many times greater than that for electrons. 
It is then reasonable to assume that electrons immediately adjust to changes in nuclear positions, 
which allows for electronic and nuclear degrees of freedom to be handled separately in the Schrödinger 
equation -- this is the essential content of the well-known Born-Oppenheimer approximation. Under 
this framework, the nuclear coordinates~$\{\mathbf{R}\}$ can be independently treated as classical 
variables that parametrize an effective Hamiltonian (the operator within square brackets in the 
left-hand side of equation~\eqref{eq.electronic.problem} below) which acts on electron coordinates 
$\{\mbf{r}\}$ only -- thus, the molecular problem amounts to finding wave functions 
$\Psi(\mbf{r}_1, \dots, \mbf{r}_{N_e}; \{\mbf{R}\})$ satisfying
\begin{equation}\label{eq.electronic.problem}
\left[ -\sum^{N_e}_{i=1}\frac{\hbar^2 \nabla^2_i}{2m_e}
-\sum^{N_e}_{i=1} \sum^{N_n}_{A=1} \frac{k_0 Z_A e^2}{|\mbf{r}_i-\mbf{R}_A|}
+\sum^{N_e}_{i=1} \sum_{j>i} \frac{k_0 e^2}{|\mbf{r}_i-\mbf{r}_j|}
+U_{nn}\{\mbf{R}\}
\right] \Psi = E(\mbf{R}_1,\dots,\mbf{R}_{N_n}) \Psi
\end{equation}
Here, $\hbar$ is the reduced Planck constant, $m_e$ the electron mass, $e$ the elementary charge, $k_0$ Coulomb's
constant, and $Z_A$ is the (effective) atomic number of $A$-th nucleus. The term $U_{nn}\{\mbf{R}\}$ is the
electrostatic repulsion between nuclei and does not depend on electronic coordinates. The energy levels obtained 
by solving the eigenproblem posed by equation~\eqref{eq.electronic.problem} are functions of the nuclear
positions, varying as the molecular geometry changes -- each level constitutes what is known as a 
\textit{potential energy surface}, an invaluable concept for understanding many chemical processes 
(electronic transitions, structural changes, etc.). The surface formed by selecting the lowest 
energy value at each nuclear configuration is of particular interest, for it describes the normal 
state of the molecule: its gradients determine the intramolecular forces and its global minimum 
defines the equilibrium geometry.\footnote{When more than one potential energy surface is involved 
in the description, one enters the realm of \textit{non-adiabatic dynamics} -- in photochemistry, 
for example, laser-induced nuclear vibrations are capable of promoting transitions among otherwise 
adiabatic energy surfaces.} Information concerning the molecule's electron charge distribution is 
built into this energy expression since the latter is an eigenvalue of the electronic wave function.

\subsection{Solving the molecular electronic structure}
\label{sub:hf.dft}

Solving the eigenvalue problem~\eqref{eq.electronic.problem} by standard diagonalisation
techniques is not a viable option in a molecular context, except for very small systems,
and one must resort to approximate schemes. The most popular and well-established methods
employed for this task are the Hartree-Fock approximation\cite{slater1963quantum} (HF) and
density functional theory\cite{hohenberg1964inhomogeneous, kohn1965self} (DFT), their success
owing to their overall balance between accuracy and computational cost. Despite their being
founded on different theoretical grounds, both methods are based on the strategy of reducing
the many-electron problem to a set of effective single-particle equations that have to be
solved self-consistently.

The HF approach is based on the idea that each electron of the system can be approximately
described as an independent particle experiencing an average external field due to the
presence of the other electrons and nuclei. It outputs optimized single-particle molecular orbitals
thus providing an intuitive understanding of electronic processes. The need to evaluate non-local
(high-cost) exchange integrals constitutes its major drawback. Meanwhile, DFT rests on the fundamental
Hohenberg-Kohn theorems which, in few words, state that the lowest energy of a many-electron system
can be obtained through a variational procedure where the electronic density -- a very concrete
observable (as opposed to the many-body wave function) -- is taken as the basic variable. DFT
is an exact theory in principle but only approximate in practice because the precise form of
the universal energy functional, the central object of the theory, is not known.\cite{capelle2006bird}
It is perhaps not as interpretive as HF but it can be formulated in terms of local (low-cost) integrals
only, being more efficient than HF in most cases. It has become routine to compute, using DFT, the
electronic structure of molecules containing hundreds of atoms, with overall good agreement between
numerical results and experimental measurements. State-of-the-art algorithms are now reaching the
thousand-atom regime.\cite{Ratcliff2016}

\subsection{The many-molecule problem and the need for force field techniques}
\label{sub:many.molecule.problem}

Ideally, one would run an atomistic simulation by solving, at each time step, the electronic
structure of the entire system in order to get the required forces: indeed, the formulation
so far outlined for isolated molecules can be directly extended to a collection of molecules,
in which case the output potential energy surface would be a function of the coordinates of
\textit{all atoms} contained in the simulation box. The envisioned function would then determine, 
through equation~\eqref{eq.force.as.gradient}, both intra- and intermolecular forces at a fixed 
time instant -- the resulting framework would constitute what is called an ab initio molecular-dynamics 
approach. In the simplest applications, however, one typically has tens of thousands of atoms in the 
simulation box, and the equations of motion need to be integrated for thousands of time steps. And, 
while it is still possible to handle the electronic problem for a snapshot of the system, it is simply 
not affordable to do molecular dynamics in this way, at least not for the sizes (number of atoms) and 
time spans needed for reliable statistics. Force field methods are precisely designed to overcome this 
difficulty, and they basically follow two types of strategies, which we may call `perturbative approach' 
and `parameter-fitting approach'.

In a perturbative approach, the energy expression is derived from quantum mechanics by a systematic
simplification of the many-molecule problem.\cite{Xu2018} One example of this class of techniques
is the Effective Fragment Potential (EFP) method.\cite{Gordon2001,Jensen2001,Gordon2013} The essential
idea of EFP is to write the interaction energy between molecules, or `fragments', using a small set of
low-cost intermolecular terms together with pre-calculated molecular orbitals from each individual fragment
obtained as if they were isolated. The approach is affordable because this last input can be independently
computed for each molecule in the system. The method operates at the Hartree-Fock level of theory, where
the energy of two nearby fragments depends on a set of parameters that control how the valence electrons
of each species redistribute themselves in order to minimize the total energy of the fragment pair. At
moderate intermolecular range these parameters are presumably small -- one is then able to identify negligible
quantities and, following a number of approximations, the mixing parameters can be efficiently computed. In
this way, a tractable expression is obtained for the pair's interaction energy, which is later extended to
a many-molecule context, although this is not straightforward due to the non-additive character of some of
the energy terms (giving rise to so-called many-body effects).\cite{stone2013theory} The resulting formulas 
in EFP are written in terms of quantum ingredients and the various energy contributions -- exchange-repulsion,
induction, charge transfer, etc. -- can be associated with distinct electronic mechanisms. Among other things, 
this means that EFP can be conveniently employed to interface a molecular dynamics domain (solvent) with a 
chemically active region (solute), the latter receiving a full quantum treatment -- the method was in fact 
originally intended for this type of application.

In a parameter-fitting approach, on the other hand, the difficulties stemming from the complexity
of the quantum problem are circumvented by adopting a completely different strategy. Here, one relies 
on physical intuition to write the energy as a sum of contributions whose functional dependence on 
atomic coordinates is designed to model specific types of interactions in an attempt to emulate the 
behavior that would follow from a complete quantum description. Usually, bonded (intramolecular)
interactions are set to describe simple classical motion -- stretching, bending, torsion, vibration,
etc. -- whereas intermolecular terms involve electrostatic and van der Waals interactions, often
including correction factors in order to account for charge-screening and other effects. Each 
energy term depends on a set of adjustable parameters chosen so that selected features of the 
substance are correctly reproduced. This optimization can be made with respect to experimental 
measurements of bulk properties and transport coefficients, or with respect to numerical results 
from a more sophisticated calculation -- for instance, a short-time few-molecule DFT-assisted 
molecular dynamics simulation.\cite{Tangney2002, Tangney2003} In this latter scheme, a more 
detailed fitting is possible, since one may resort to experimentally inaccessible information, 
such as stress components or molecular forces, to carry out the optimization; indeed, it is 
a popular trend nowadays to employ machine learning algorithms for this task.\cite{rupp2018guest} 
The main limitation of this kind of approach is that parameters are necessarily fitted to a 
predetermined thermodynamic state and thus one should not expect the force field to be accurate 
under different conditions -- the so-called \textit{transferability} issue. Nevertheless, a 
well-designed force field can retain its accuracy for a sufficiently broad range of thermodynamic 
variables and, if this is the case, it will most likely outperform other approaches in practical 
applications.


\section{Final remarks}
\label{sec:final.remarks}

By considering the particular problem of determining material-specific boundary conditions at fluid-solid 
interfaces, two transitions of space-time scale were discussed. First, by introducing the concept of slip 
length and, thus, establishing a third-type boundary condition, a methodology for the effective application 
of molecular-dynamics results in the context of macroscopic flows was delineated. In this way, unlike the 
situation prescribed by the no-slip condition, the dynamics of fluids constrained by solid surfaces acquires 
a description that depends not only on the chemical composition of the interfacial materials, but also on 
their thermodynamic states, as specified by field variables such as temperature and pressure, and microscopic 
spatial configuration, which influences physical characteristics such as surface patterning and roughness. 
The ability to perform material-specific simulations of macroscopic flows provides a very refined analytical 
tool, which can assist the design, development, and optimization of technologically relevant devices.

In order to determine the interatomic force fields required for the accurate description of a particular 
material in a molecular dynamics simulation, a second scale transition was also examined. In this case,
brief considerations were made on the employment of numerical methods of quantum mechanics in obtaining 
effective classical potentials of intra- and intermolecular interaction. In particular, two strategies 
were addressed: a perturbative approach, which relies on successive simplifications of the complete 
quantum problem of many interacting particles, and a parametric fitting of potential energy functions, 
which depends on the comparison with empirical observations or extensive quantum computations.

The multiscale approaches presented in this elementary overview are based on the fact 
that calculations on different spatial scales can be done \textit{independently}, that 
is, physical properties can be evaluated on a finer scale, by considering more fundamental 
theories of matter, and \textit{subsequently} employed in solving problems on a larger 
scale, where insufficient theoretical details were initially available. This procedure 
is only possible because of the great difference also existing on the time scales 
characterizing distinct physical phenomena, so that the dynamics of a system and 
its small-sized subsystems can be treated as effectively decoupled.


\section*{Acknowledgments}

This work is part of a project developed in the Research Centre for Gas Innovation (RCGI),
with support from Shell and FAPESP (Fundação de Amparo à Pesquisa do Estado de São Paulo), 
under process numbers 2014/50279-4 and 2018/03211-6.

\bibliographystyle{unsrt}
\bibliography{BoundaryConditions,ForceFields} 

\begin{thebibliography}{100}

\bibitem{landau1987fluid}
L.~D. Landau and E.~M. Lifshitz.
\newblock {\em Fluid mechanics}.
\newblock Pergamon Press, 1987.

\bibitem{grad1949kinetic}
Harold Grad.
\newblock On the kinetic theory of rarefied gases.
\newblock {\em Communications on Pure and Applied Mathematics}, 2(4):331--407,
  1949.

\bibitem{Chapman1990}
Sydney Chapman, Thomas~George Cowling, and David Burnett.
\newblock {\em The mathematical theory of non-uniform gases: an account of the
  kinetic theory of viscosity, thermal conduction and diffusion in gases}.
\newblock Cambridge university press, 1990.

\bibitem{Kremer2005}
Gilberto~Medeiros Kremer.
\newblock {\em Uma Introdu{\c{c}}{\~a}o {\`a} Equa{\c{c}}{\~a}o de Boltzmann}.
\newblock Edusp, 2005.

\bibitem{irving1950statistical}
J.~H. Irving and John~G. Kirkwood.
\newblock The statistical mechanical theory of transport processes. {IV}. the
  equations of hydrodynamics.
\newblock {\em The Journal of Chemical Physics}, 18(6):817--829, 1950.

\bibitem{Asproulis2012}
Nikolaos Asproulis, Marco Kalweit, and Dimitris Drikakis.
\newblock {A hybrid molecular continuum method using point wise coupling}.
\newblock {\em Advances in Engineering Software}, 46(1):85--92, 2012.

\bibitem{Shang2012a}
Barry~Z. Shang, Nikolaos~K. Voulgarakis, and Jhih-Wei Chu.
\newblock {Fluctuating hydrodynamics for multiscale modeling and simulation:
  Energy and heat transfer in molecular fluids}.
\newblock {\em The Journal of Chemical Physics}, 137(4):044117, jul 2012.

\bibitem{Borg2013a}
Matthew~K. Borg, Duncan~A. Lockerby, and Jason~M. Reese.
\newblock {A hybrid molecular-continuum simulation method for incompressible
  flows in micro/nanofluidic networks}.
\newblock {\em Microfluidics and Nanofluidics}, 15(4):541--557, 2013.

\bibitem{Cosden2013}
Ian~A. Cosden and Jennifer~R. Lukes.
\newblock {A hybrid atomistic-continuum model for fluid flow using LAMMPS and
  OpenFOAM}.
\newblock {\em Computer Physics Communications}, 184(8):1958--1965, 2013.

\bibitem{Markesteijn2014}
Anton Markesteijn, Sergey Karabasov, Arturs Scukins, Dmitry Nerukh, Vyacheslav
  Glotov, and Vasily Goloviznin.
\newblock {Concurrent multiscale modelling of atomistic and hydrodynamic
  processes in liquids}.
\newblock {\em Philosophical Transactions of the Royal Society A: Mathematical,
  Physical and Engineering Sciences}, 372(2021):20130379--20130379, jun 2014.

\bibitem{allen2017computer}
Michael~P. Allen and Dominic~J. Tildesley.
\newblock {\em Computer simulation of liquids}.
\newblock Oxford university press, 2017.

\bibitem{hoover1985canonical}
William~G. Hoover.
\newblock Canonical dynamics: equilibrium phase-space distributions.
\newblock {\em Physical review A}, 31(3):1695, 1985.

\bibitem{hoover1986constant}
William~G. Hoover.
\newblock Constant-pressure equations of motion.
\newblock {\em Physical Review A}, 34(3):2499, 1986.

\bibitem{rapaport2004art}
Dennis~C. Rapaport.
\newblock {\em The Art of Molecular Dynamics Simulation}.
\newblock Cambridge university press, 2004.

\bibitem{edberg1987rheology}
Roger Edberg, G.~P. Morriss, and Denis~J. Evans.
\newblock Rheology of n-alkanes by nonequilibrium molecular dynamics.
\newblock {\em The Journal of Chemical Physics}, 86(8):4555--4570, 1987.

\bibitem{morriss2007statistical}
Gary~P. Morriss and Denis~J. Evans.
\newblock {\em Statistical Mechanics of Nonequilbrium Liquids}.
\newblock ANU Press, 2007.

\bibitem{day1990no}
Michael~A. Day.
\newblock The no-slip condition of fluid dynamics.
\newblock {\em Erkenntnis}, 33(3):285--296, 1990.

\bibitem{lauga2007microfluidics}
Eric Lauga, Michael Brenner, and Howard Stone.
\newblock Microfluidics: the no-slip boundary condition.
\newblock In {\em Springer Handbook of Experimental Fluid Mechanics}, pages
  1219--1240. Springer, 2007.

\bibitem{granick2003slippery}
Steve Granick, Yingxi Zhu, and Hyunjung Lee.
\newblock Slippery questions about complex fluids flowing past solids.
\newblock {\em Nature Materials}, 2(4):221, 2003.

\bibitem{ellis2004slip}
Jonathan~S. Ellis and Michael Thompson.
\newblock Slip and coupling phenomena at the liquid--solid interface.
\newblock {\em Physical Chemistry Chemical Physics}, 6(21):4928--4938, 2004.

\bibitem{neto2005boundary}
Chiara Neto, Drew~R. Evans, Elmar Bonaccurso, Hans-J{\"u}rgen Butt, and Vincent
  S.~J. Craig.
\newblock Boundary slip in newtonian liquids: a review of experimental studies.
\newblock {\em Reports on Progress in Physics}, 68(12):2859, 2005.

\bibitem{thompson1989simulations}
Peter~A. Thompson and Mark~O. Robbins.
\newblock Simulations of contact-line motion: slip and the dynamic contact
  angle.
\newblock {\em Physical Review Letters}, 63(7):766, 1989.

\bibitem{koplik1998no}
Joel Koplik and Jayanth~R. Banavar.
\newblock No-slip condition for a mixture of two liquids.
\newblock {\em Physical Review Letters}, 80(23):5125, 1998.

\bibitem{sokhan2001fluid}
V.~P. Sokhan, D.~Nicholson, and N.~Quirke.
\newblock Fluid flow in nanopores: an examination of hydrodynamic boundary
  conditions.
\newblock {\em The Journal of Chemical Physics}, 115(8):3878--3887, 2001.

\bibitem{sokhan2002fluid}
Vladimir~P. Sokhan, David Nicholson, and Nicholas Quirke.
\newblock Fluid flow in nanopores: Accurate boundary conditions for carbon
  nanotubes.
\newblock {\em The Journal of Chemical Physics}, 117(18):8531--8539, 2002.

\bibitem{koplik1989molecular}
Joel Koplik, Jayanth~R. Banavar, and Jorge~F. Willemsen.
\newblock Molecular dynamics of fluid flow at solid surfaces.
\newblock {\em Physics of Fluids A: Fluid Dynamics}, 1(5):781--794, 1989.

\bibitem{cieplak2001boundary}
Marek Cieplak, Joel Koplik, and Jayanth~R. Banavar.
\newblock Boundary conditions at a fluid-solid interface.
\newblock {\em Physical Review Letters}, 86(5):803, 2001.

\bibitem{gupta1997shear}
S.~A. Gupta, H.~D. Cochran, and P.~T. Cummings.
\newblock Shear behavior of squalane and tetracosane under extreme confinement.
  i. model, simulation method, and interfacial slip.
\newblock {\em The Journal of Chemical Physics}, 107(23):10316--10326, 1997.

\bibitem{jabbarzadeh1999wall}
A.~Jabbarzadeh, J.~D.Atkinson, and R.~I. Tanner.
\newblock Wall slip in the molecular dynamics simulation of thin films of
  hexadecane.
\newblock {\em The Journal of Chemical Physics}, 110(5):2612--2620, 1999.

\bibitem{liu2011validity}
Chong Liu and Zhigang Li.
\newblock On the validity of the navier-stokes equations for nanoscale liquid
  flows: The role of channel size.
\newblock {\em AIP Advances}, 1(3):032108, 2011.

\bibitem{xu2007boundary}
Jinliang Xu and Yuxiu Li.
\newblock Boundary conditions at the solid--liquid surface over the multiscale
  channel size from nanometer to micron.
\newblock {\em International Journal of Heat and Mass Transfer},
  50(13-14):2571--2581, 2007.

\bibitem{thomas2008reassessing}
John~A. Thomas and Alan J.~H. McGaughey.
\newblock Reassessing fast water transport through carbon nanotubes.
\newblock {\em Nano Letters}, 8(9):2788--2793, 2008.

\bibitem{ramos2016hydrodynamic}
Bladimir Ramos-Alvarado, Satish Kumar, and G.~P. Peterson.
\newblock Hydrodynamic slip length as a surface property.
\newblock {\em Physical Review E}, 93(2):023101, 2016.

\bibitem{thompson1997general}
Peter~A. Thompson and Sandra~M. Troian.
\newblock A general boundary condition for liquid flow at solid surfaces.
\newblock {\em Nature}, 389(6649):360, 1997.

\bibitem{priezjev2004molecular}
Nikolai~V. Priezjev and Sandra~M. Troian.
\newblock Molecular origin and dynamic behavior of slip in sheared polymer
  films.
\newblock {\em Physical Review Letters}, 92(1):018302, 2004.

\bibitem{pahlavan2011effect}
Amir~Alizadeh Pahlavan and Jonathan~B. Freund.
\newblock Effect of solid properties on slip at a fluid-solid interface.
\newblock {\em Physical Review E}, 83(2):021602, 2011.

\bibitem{navier1823memoire}
C.~L. M.~H. Navier.
\newblock M{\'e}moire sur les lois du mouvement des fluides.
\newblock {\em M{\'e}moires de l'Acad{\'e}mie Royale des Sciences de l'Institut
  de France}, 6(1823):389--440, 1823.

\bibitem{khare1996rheology}
Rajesh Khare, Juan~J. de~Pablo, and Arun Yethiraj.
\newblock Rheology of confined polymer melts.
\newblock {\em Macromolecules}, 29(24):7910--7918, 1996.

\bibitem{jabbarzadeh2002effect}
A.~Jabbarzadeh, J.~D. Atkinson, and R.~I. Tanner.
\newblock The effect of branching on slip and rheological properties of
  lubricants in molecular dynamics simulation of couette shear flow.
\newblock {\em Tribology International}, 35(1):35--46, 2002.

\bibitem{niavarani2008rheological}
Anoosheh Niavarani and Nikolai~V. Priezjev.
\newblock Rheological study of polymer flow past rough surfaces with slip
  boundary conditions.
\newblock {\em The Journal of Chemical Physics}, 129(14):144902, 2008.

\bibitem{niavarani2008slip}
Anoosheh Niavarani and Nikolai~V. Priezjev.
\newblock Slip boundary conditions for shear flow of polymer melts past
  atomically flat surfaces.
\newblock {\em Physical Review E}, 77(4):041606, 2008.

\bibitem{priezjev2009shear}
Nikolai~V. Priezjev.
\newblock Shear rate threshold for the boundary slip in dense polymer films.
\newblock {\em Physical Review E}, 80(3):031608, 2009.

\bibitem{priezjev2010relationship}
Nikolai~V. Priezjev.
\newblock Relationship between induced fluid structure and boundary slip in
  nanoscale polymer films.
\newblock {\em Physical Review E}, 82(5):051603, 2010.

\bibitem{priezjev2012interfacial}
Nikolai~V. Priezjev.
\newblock Interfacial friction between semiflexible polymers and crystalline
  surfaces.
\newblock {\em The Journal of chemical physics}, 136(22):224702, 2012.

\bibitem{priezjev2012fluid}
Nikolai~V. Priezjev.
\newblock Fluid structure and boundary slippage in nanoscale liquid films.
\newblock {\em Detection of Pathogens in Water Using Micro and
  Nano-technology}, 2012.

\bibitem{priezjev2007rate}
Nikolai~V. Priezjev.
\newblock Rate-dependent slip boundary conditions for simple fluids.
\newblock {\em Physical Review E}, 75(5):051605, 2007.

\bibitem{martini2008slip}
Ashlie Martini, Hua-Yi Hsu, Neelesh~A. Patankar, and Seth Lichter.
\newblock Slip at high shear rates.
\newblock {\em Physical Review Letters}, 100(20):206001, 2008.

\bibitem{asproulis2010boundary}
Nikolaos Asproulis and Dimitris Drikakis.
\newblock Boundary slip dependency on surface stiffness.
\newblock {\em Physical Review E}, 81(6):061503, 2010.

\bibitem{asproulis2011wall}
Nikolaos Asproulis and Dimitris Drikakis.
\newblock Wall-mass effects on hydrodynamic boundary slip.
\newblock {\em Physical Review E}, 84(3):031504, 2011.

\bibitem{hunenberger2005thermostat}
Philippe~H. H{\"u}nenberger.
\newblock Thermostat algorithms for molecular dynamics simulations.
\newblock In {\em Advanced Computer Simulation}, pages 105--149. Springer,
  2005.

\bibitem{liu2009flow}
Chong Liu and Zhigang Li.
\newblock Flow regimes and parameter dependence in nanochannel flows.
\newblock {\em Physical Review E}, 80(3):036302, 2009.

\bibitem{liu2010surface}
Chong Liu and Zhigang Li.
\newblock Surface effects on nanoscale poiseuille flows under large driving
  force.
\newblock {\em The Journal of Chemical Physics}, 132(2):024507, 2010.

\bibitem{semiromi2010nanoscale}
D.~Toghraie Semiromi and A.~R. Azimian.
\newblock Nanoscale poiseuille flow and effects of modified lennard--jones
  potential function.
\newblock {\em Heat and Mass Transfer}, 46(7):791--801, 2010.

\bibitem{sofos2013parameters}
Filippos Sofos, Theodoros~E. Karakasidis, and Antonios Liakopoulos.
\newblock Parameters affecting slip length at the nanoscale.
\newblock {\em Journal of Computational and Theoretical Nanoscience},
  10(3):648--650, 2013.

\bibitem{jabbarzadeh2000effect}
A.~Jabbarzadeh, J.~D. Atkinson, and R.~I. Tanner.
\newblock Effect of the wall roughness on slip and rheological properties of
  hexadecane in molecular dynamics simulation of couette shear flow between two
  sinusoidal walls.
\newblock {\em Physical Review E}, 61(1):690, 2000.

\bibitem{cottin2003low}
C{\'e}cile Cottin-Bizonne, Jean-Louis Barrat, Lyd{\'e}ric Bocquet, and
  Elisabeth Charlaix.
\newblock Low-friction flows of liquid at nanopatterned interfaces.
\newblock {\em Nature Materials}, 2(4):237, 2003.

\bibitem{cottin2004dynamics}
C{\'e}cile Cottin-Bizonne, Catherine Barentin, {\'E}lisabeth Charlaix,
  Lyd{\'e}ric Bocquet, and J.-L. Barrat.
\newblock Dynamics of simple liquids at heterogeneous surfaces:
  Molecular-dynamics simulations and hydrodynamic description.
\newblock {\em The European Physical Journal E}, 15(4):427--438, 2004.

\bibitem{galea2004molecular}
Toni-Marie Galea and Phil Attard.
\newblock Molecular dynamics study of the effect of atomic roughness on the
  slip length at the fluid- solid boundary during shear flow.
\newblock {\em Langmuir}, 20(8):3477--3482, 2004.

\bibitem{priezjev2005slip}
Nikolai~V. Priezjev, Anton~A. Darhuber, and Sandra~M. Troian.
\newblock Slip behavior in liquid films on surfaces of patterned wettability:
  Comparison between continuum and molecular dynamics simulations.
\newblock {\em Physical Review E}, 71(4):041608, 2005.

\bibitem{priezjev2006influence}
Nikolai~V. Priezjev and Sandra~M. Troian.
\newblock Influence of periodic wall roughness on the slip behaviour at
  liquid/solid interfaces: molecular-scale simulations versus continuum
  predictions.
\newblock {\em Journal of Fluid Mechanics}, 554:25--46, 2006.

\bibitem{priezjev2007effect}
Nikolai~V. Priezjev.
\newblock Effect of surface roughness on rate-dependent slip in simple fluids.
\newblock {\em The Journal of Chemical Physics}, 127(14):144708, 2007.

\bibitem{niavarani2010modeling}
Anoosheh Niavarani and Nikolai~V. Priezjev.
\newblock Modeling the combined effect of surface roughness and shear rate on
  slip flow of simple fluids.
\newblock {\em Physical Review E}, 81(1):011606, 2010.

\bibitem{guo2005temperature}
Zhaoli Guo, T.~S. Zhao, and Yong Shi.
\newblock Temperature dependence of the velocity boundary condition for
  nanoscale fluid flows.
\newblock {\em Physical Review E}, 72(3):036301, 2005.

\bibitem{servantie2008temperature}
J.~Servantie and M.~M{\"u}ller.
\newblock Temperature dependence of the slip length in polymer melts at
  attractive surfaces.
\newblock {\em Physical Review Letters}, 101(2):026101, 2008.

\bibitem{bao2017effects}
Luyao Bao, Nikolai~V. Priezjev, Haibao Hu, and Kai Luo.
\newblock Effects of viscous heating and wall-fluid interaction energy on
  rate-dependent slip behavior of simple fluids.
\newblock {\em Physical Review E}, 96(3):033110, 2017.

\bibitem{lichter2004mechanisms}
Seth Lichter, Alex Roxin, and Shreyas Mandre.
\newblock Mechanisms for liquid slip at solid surfaces.
\newblock {\em Physical Review Letters}, 93(8):086001, 2004.

\bibitem{lichter2007liquid}
Seth Lichter, Ashlie Martini, Randall~Q. Snurr, and Qian Wang.
\newblock Liquid slip in nanoscale channels as a rate process.
\newblock {\em Physical Review Letters}, 98(22):226001, 2007.

\bibitem{martini2008molecular}
A.~Martini, A.~Roxin, R.~Q. Snurr, Q.~Wang, and S.~Lichter.
\newblock Molecular mechanisms of liquid slip.
\newblock {\em Journal of Fluid Mechanics}, 600:257--269, 2008.

\bibitem{yong2010investigating}
Xin Yong and Lucy~T. Zhang.
\newblock Investigating liquid-solid interfacial phenomena in a couette flow at
  nanoscale.
\newblock {\em Physical Review E}, 82(5):056313, 2010.

\bibitem{sochi2011slip}
Taha Sochi.
\newblock Slip at fluid-solid interface.
\newblock {\em Polymer Reviews}, 51(4):309--340, 2011.

\bibitem{yong2013slip}
Xin Yong and Lucy~T. Zhang.
\newblock Slip in nanoscale shear flow: mechanisms of interfacial friction.
\newblock {\em Microfluidics and Nanofluidics}, 14(1-2):299--308, 2013.

\bibitem{barrat1999large}
Jean-Louis Barrat and Lyd{\'e}ric Bocquet.
\newblock Large slip effect at a nonwetting fluid-solid interface.
\newblock {\em Physical Review Letters}, 82(23):4671, 1999.

\bibitem{barrat1999influence}
Jean-Louis Barrat and Lyd{\'e}ric Bocquet.
\newblock Influence of wetting properties on hydrodynamic boundary conditions
  at a fluid/solid interface.
\newblock {\em Faraday Discussions}, 112:119--128, 1999.

\bibitem{nagayama2004effects}
Gyoko Nagayama and Ping Cheng.
\newblock Effects of interface wettability on microscale flow by molecular
  dynamics simulation.
\newblock {\em International Journal of Heat and Mass Transfer},
  47(3):501--513, 2004.

\bibitem{huang2008water}
David~M. Huang, Christian Sendner, Dominik Horinek, Roland~R. Netz, and
  Lyd{\'e}ric Bocquet.
\newblock Water slippage versus contact angle: A quasiuniversal relationship.
\newblock {\em Physical Review Letters}, 101(22):226101, 2008.

\bibitem{sendner2009interfacial}
Christian Sendner, Dominik Horinek, Lyderic Bocquet, and Roland~R. Netz.
\newblock Interfacial water at hydrophobic and hydrophilic surfaces: Slip,
  viscosity, and diffusion.
\newblock {\em Langmuir}, 25(18):10768--10781, 2009.

\bibitem{huang2012friction}
Kai Huang and Izabela Szlufarska.
\newblock Friction and slip at the solid/liquid interface in vibrational
  systems.
\newblock {\em Langmuir}, 28(50):17302--17312, 2012.

\bibitem{ramos2016wettability}
Bladimir Ramos-Alvarado, Satish Kumar, and G.~P. Peterson.
\newblock Wettability transparency and the quasiuniversal relationship between
  hydrodynamic slip and contact angle.
\newblock {\em Applied Physics Letters}, 108(7):074105, 2016.

\bibitem{yen2016effective}
Tsu-Hsu Yen and Chyi-Yeou Soong.
\newblock Effective boundary slip and wetting characteristics of water on
  substrates with effects of surface morphology.
\newblock {\em Molecular Physics}, 114(6):797--809, 2016.

\bibitem{voronov2006boundary}
Roman~S. Voronov, Dimitrios~V. Papavassiliou, and Lloyd~L. Lee.
\newblock Boundary slip and wetting properties of interfaces: Correlation of
  the contact angle with the slip length.
\newblock {\em The Journal of Chemical Physics}, 124(20):204701, 2006.

\bibitem{voronov2007slip}
Roman~S. Voronov, Dimitrios~V. Papavassiliou, and Lloyd~L. Lee.
\newblock Slip length and contact angle over hydrophobic surfaces.
\newblock {\em Chemical Physics Letters}, 441(4-6):273--276, 2007.

\bibitem{voronov2008review}
Roman~S. Voronov, Dimitrios~V. Papavassiliou, and Lloyd~L. Lee.
\newblock Review of fluid slip over superhydrophobic surfaces and its
  dependence on the contact angle.
\newblock {\em Industrial \& Engineering Chemistry Research}, 47(8):2455--2477,
  2008.

\bibitem{ho2011liquid}
Tuan~Anh Ho, Dimitrios~V. Papavassiliou, Lloyd~L. Lee, and Alberto Striolo.
\newblock Liquid water can slip on a hydrophilic surface.
\newblock {\em Proceedings of the National Academy of Sciences}, 2011.

\bibitem{brooks1983charmm}
Bernard~R. Brooks, Robert~E. Bruccoleri, Barry~D. Olafson, David~J. States,
  S.~Swaminathan, and Martin Karplus.
\newblock Charmm: a program for macromolecular energy, minimization, and
  dynamics calculations.
\newblock {\em Journal of Computational Chemistry}, 4(2):187--217, 1983.

\bibitem{daw1984embedded}
Murray~S. Daw and Michael~I. Baskes.
\newblock Embedded-atom method: Derivation and application to impurities,
  surfaces, and other defects in metals.
\newblock {\em Physical Review B}, 29(12):6443, 1984.

\bibitem{van2001reaxff}
Adri C.~T. van Duin, Siddharth Dasgupta, Francois Lorant, and William~A.
  Goddard.
\newblock Reaxff: a reactive force field for hydrocarbons.
\newblock {\em The Journal of Physical Chemistry A}, 105(41):9396--9409, 2001.

\bibitem{ponder2003force}
Jay~W. Ponder and David~A. Case.
\newblock Force fields for protein simulations.
\newblock In {\em Advances in Protein Chemistry}, volume~66, pages 27--85.
  Elsevier, 2003.

\bibitem{lopes2015current}
Pedro E.~M. Lopes, Olgun Guvench, and Alexander~D. MacKerell.
\newblock Current status of protein force fields for molecular dynamics
  simulations.
\newblock In {\em Molecular Modeling of Proteins}, pages 47--71. Springer,
  2015.

\bibitem{atkins2011molecular}
Peter~W. Atkins and Ronald~S. Friedman.
\newblock {\em Molecular quantum mechanics}.
\newblock Oxford university press, 2011.

\bibitem{szabo2012modern}
Attila Szabo and Neil~S. Ostlund.
\newblock {\em Modern quantum chemistry: introduction to advanced electronic
  structure theory}.
\newblock Courier Corporation, 2012.

\bibitem{slater1963quantum}
John~Clarke Slater.
\newblock {\em Quantum theory of molecules and solids}, volume~1.
\newblock McGraw-Hill New York, 1963.

\bibitem{hohenberg1964inhomogeneous}
Pierre Hohenberg and Walter Kohn.
\newblock Inhomogeneous electron gas.
\newblock {\em Physical Review}, 136(3B):B864, 1964.

\bibitem{kohn1965self}
Walter Kohn and Lu~Jeu Sham.
\newblock Self-consistent equations including exchange and correlation effects.
\newblock {\em Physical review}, 140(4A):A1133, 1965.

\bibitem{capelle2006bird}
Klaus Capelle.
\newblock A bird's-eye view of density-functional theory.
\newblock {\em Brazilian Journal of Physics}, 36(4A):1318--1343, 2006.

\bibitem{Ratcliff2016}
Laura~E. Ratcliff, Stephan Mohr, Georg Huhs, Thierry Deutsch, Michel Masella,
  and Luigi Genovese.
\newblock Challenges in large scale quantum mechanical calculations.
\newblock {\em Wiley Interdisciplinary Reviews: Computational Molecular
  Science}, 7(1):e1290, 2017.

\bibitem{Xu2018}
Peng Xu, Emilie~B. Guidez, Colleen Bertoni, and Mark~S. Gordon.
\newblock {Perspective: Ab initio force field methods derived from quantum
  mechanics}.
\newblock {\em Journal of Chemical Physics}, 148(9):090901, 2018.

\bibitem{Gordon2001}
Mark~S. Gordon, Mark~A. Freitag, Pradipta Bandyopadhyay, Jan~H. Jensen,
  Visvaldas Kairys, and Walter~J. Stevens.
\newblock {The Effective Fragment Potential Method:  A QM-Based MM Approach to
  Modeling Environmental Effects in Chemistry}.
\newblock {\em The Journal of Physical Chemistry A}, 105(2):293--307, 2001.

\bibitem{Jensen2001}
Jan~H. Jensen.
\newblock {Intermolecular exchange-induction and charge transfer: Derivation of
  approximate formulas using nonorthogonal localized molecular orbitals}.
\newblock {\em The Journal of Chemical Physics}, 114(20):8775--8783, may 2001.

\bibitem{Gordon2013}
Mark~S. Gordon, Quentin~A. Smith, Peng Xu, and Lyudmila~V. Slipchenko.
\newblock {Accurate First Principles Model Potentials for Intermolecular
  Interactions}.
\newblock {\em Annual Review of Physical Chemistry}, 64(1):553--578, 2013.

\bibitem{stone2013theory}
Anthony Stone.
\newblock {\em The theory of intermolecular forces}.
\newblock OUP Oxford, 2013.

\bibitem{Tangney2002}
P.~Tangney and S.~Scandolo.
\newblock {An ab initio parametrized interatomic force field for silica}.
\newblock {\em The Journal of Chemical Physics}, 117(19):8898--8904, nov 2002.

\bibitem{Tangney2003}
R.~Tangney and S.~Scandolo.
\newblock {A many-body interatomic potential for ionic systems: Application to
  MgO}.
\newblock {\em Journal of Chemical Physics}, 119(18):9673--9685, 2003.

\bibitem{rupp2018guest}
Matthias Rupp, O.~Anatole von Lilienfeld, and Kieron Burke.
\newblock Guest editorial: Special topic on data-enabled theoretical chemistry,
  2018.

\end{thebibliography}

\end{document}